\documentclass[aps,prc,a4paper,nofootinbib,
preprintnumbers,superscriptaddress,showpacs,showkeys,twocolumn]{revtex4}

\usepackage{amsmath}
\usepackage{amssymb}
\usepackage{bm}
\usepackage{graphicx}
\usepackage{xcolor}

\newcommand{\be}{\begin{equation}}
\newcommand{\ee}{\end{equation}}
\newcommand{\ba}{\begin{eqnarray}}
\newcommand{\ea}{\end{eqnarray}}

\begin{document}

\title{Ballistic diffusion of heavy quarks in the early stage of relativistic heavy ion collisions at RHIC and LHC}

\author{Jun-Hong Liu}
\affiliation{School of Nuclear Science and Technology, Lanzhou University, 222 South Tianshui Road, Lanzhou 730000, China}

\author{Santosh K. Das}
\affiliation{School of Physical Science, Indian Institute of Technology Goa, Ponda-403401, Goa, India}

\author{Vincenzo Greco}
\affiliation{Department of Physics and Astronomy, University of Catania, Via S. Sofia 64, I-95125 Catania}
\affiliation{INFN-Laboratori Nazionali del Sud, Via S. Sofia 62, I-95123 Catania, Italy}

\author{Marco Ruggieri}\email{ruggieri@lzu.edu.cn}
\affiliation{School of Nuclear Science and Technology, Lanzhou University, 222 South Tianshui Road, Lanzhou 730000, China}


\begin{abstract}
We study the diffusion of charm quarks in the early stage of high energy nuclear collisions
at the RHIC and the LHC. 
The main novelty of the present study is the introduction
of the  color current carried by the heavy quarks that propagate in the evolving Glasma (Ev-Glasma),
that is responsible of the energy loss via polarization of the medium. 
We compute the
transverse momentum broadening,  $\sigma_p$, of charm in the pre-thermalization stage, and the impact of the diffusion
on the nuclear modification factor in nucleus-nucleus collisions.
The net effect of energy loss is marginal in the pre-thermalization stage.  
The study is completed by the calculation of coordinate spreading, $\sigma_x$, and by a comparison with
Langevin dynamics.
 $\sigma_p$ in Ev-Glasma overshoots the result of standard Langevin dynamics at the end of the pre-hydro regime. 
We interpret this as a result of memory 
of the color force acting on the charm quarks that implies $\sigma_p\propto t^2$.  
Moreover, $\sigma_x\propto   t^2  $
in the pre-hydro stage  shows that the charm quark in the Ev-Glasma is in the regime of ballistic diffusion. 
\end{abstract}



\maketitle

\section{Introduction}
 
The study of the pre-thermalization stage of the system produced in high energy nuclear collisions  
is one of the most exciting research topics related to the physics of
relativistic heavy ion collisions (RHICs). 
Within the color-glass-condensate (CGC) effective theory 
\cite{McLerran:1993ni,McLerran:1993ka,McLerran:1994vd,Gelis:2010nm,Iancu:2003xm,McLerran:2008es,Gelis:2012ri},
the collision of two colored glasses leads to the formation of strong gluon fields in the forward light cone
named Glasma
\cite{Kovner:1995ja,Kovner:1995ts,Gyulassy:1997vt,Lappi:2006fp,Fukushima:2006ax,
Fries:2006pv,Chen:2015wia,Fujii:2008km,Krasnitz:2000gz,Krasnitz:2003jw,Krasnitz:2001qu},
consisting of longitudinal color-electric and color-magnetic fields
characterized by large gluon occupation numbers,
$A_\mu^a \simeq 1/g$ with $g$ the QCD coupling, 
so they can be described by classical field theory namely the Classical Yang-Mills  (CYM) theory.
see also \cite{Romatschke:2005pm,Romatschke:2006nk,Fukushima:2011nq,
Fukushima:2013dma,Iida:2014wea,Gelis:2013rba,Epelbaum:2013waa,Tanji:2011di,Ryblewski:2013eja,Ruggieri:2015yea,
Berges:2012cj,Berges:2013fga,Berges:2013lsa,Berges:2013eia,Ruggieri:2017ioa} for the 
next-to-leading order corrections to the Glasma.  
Once Glasma is set up as the initial condition, its evolution is studied within the CYM equations;
in this article, we denote the evolving Glasma as Ev-Glasma.

Heavy quarks, charm and beauty, are good probes of the system created in high energy nuclear collisions,
both for the pre-equilibrium part and for the thermalized quark-gluon plasma (QGP), see
\cite{Svetitsky:1987gq,Rapp:2018qla, Aarts:2016hap, Greco:2017rro,Das:2015ana, Das:2013kea,Das:2016cwd,
Das:2017dsh,Das:2015aga,Beraudo:2015wsd,Xu:2015iha,Ozvenchuk:2017ojj,
Prino:2016cni,Andronic:2015wma,
Liu:2019lac,Sun:2019fud,Ruggieri:2018rzi,Ipp:2020nfu,Boguslavski:2020tqz,
Mrowczynski:2017kso,Carrington:2020sww,Dong:2019unq,Ke:2018tsh,Yao:2020xzw} and references therein.
In fact, their formation time is $\tau_\mathrm{form} \approx 1/(2m)$ with
$m$ the quark mass which gives $\tau_\mathrm{form} \leq 0.07$ fm/c, which is shorter than the
formation time of light quarks; thus, charm and beauty propagate in the Ev-Glasma and probe its evolution.
The large mass, the early production and the low concentration of charm and beauty makes heavy quarks the perfect
probes of the medium produced in the collisions.

The evolution of heavy quarks in the Ev-Glasma has attracted a lot of interest recently,
\cite{Liu:2019lac,Sun:2019fud,Ruggieri:2018rzi,Ipp:2020nfu,Boguslavski:2020tqz,
Mrowczynski:2017kso,Carrington:2020sww}. The study of diffusion and energy loss  
was started in~\cite{Mrowczynski:2017kso}, while in \cite{Liu:2019lac,Sun:2019fud,Ruggieri:2018rzi}
it was shown for the first time how the evolution of charm and beauty in the Ev-Glasma
can affect the nuclear modification factor, $R_\mathrm{AA}$, and the elliptic flow of these quarks
both in proton-nucleus and in nucleus-nucleus collisions. More studies have been devoted to
the diffusion and momentum broadening \cite{Boguslavski:2020tqz,Carrington:2020sww,Ipp:2020nfu}. 

In previous studies, the energy loss of the heavy quarks has been neglected:
this is due to the polarization of the gluon medium induced by the color current of the heavy 
quarks \cite{Jamal:2020fxo,Han:2017nfz,Jamal:2019svc,Jamal:2020emj},
see also \cite{Ruggieri:2019zos} for a treatment of the problem within a classical model.
Adding this current results in a drag force acting on the heavy quarks: it is thus a back-reaction.
Neglecting this sounds as a reasonable approximation: in fact, the energy density of the Ev-Glasma is much larger than that
of the QGP phase, therefore the momentum broadening due to diffusion is expected to be more important
than the energy loss due to polarization. 
Nevertheless, adding the color current is a well defined procedure, therefore it is possible to add it and
compute its effect on the observables. This is one of the main purpose of the present study.

We compute the effect of the color current on the nuclear modification factor of charm in the Ev-Glasma,
on the momentum broadening and on the transverse coordinates diffusion.
This is achieved by adding this current to the CYM equations,
that we solve consistently with the kinetic equations of motion of the heavy quarks.
We find that the effect on the $R_\mathrm{AA}$ is present but not large:
this is due both to the small magnitude of the current of charm, as well as to the short lifetime
of the pre-thermalization stage. The effect of the drag force is more visible on the 
momentum and transverse plane coordinates diffusion: however, even for these quantities
the net effect of the drag is to slow down the diffusion of at most $20\%$.

We estimate both the diffusion coefficient in momentum space, ${\cal D}$,
and the drag coefficient, $\gamma$: in particular, we find that  $\gamma$ is quite small,
certainly smaller than the value it is usually used in the QGP phase.
This means that the equilibration time of the charm in the Ev-Glasma stage is much larger
than the lifetime of the Ev-Glasma itself. Thus, 
the motion of charm in the Ev-Glasma is dominated by diffusion because
the equilibration time, $\tau_\mathrm{therm}=1/\gamma$, is much larger than the lifetime
of the Ev-Glasma, as it happens for the standard Brownian motion.

We remark that this addition solves the {\em classical} problem completely and consistently:
this procedure adds the classical radiation produced by the moving heavy quarks in a consistent way,
and is qualitatively similar to what one should do in classical electrodynamics for the problem of 
the propagation of a classical electric charge in a classical electromagnetic field. In fact, 
it is well known that approaching this classical electrodynamics problem leads to the 
production of a near and a far fields, the latter being responsible of energy loss by radiation.
In solving the classical problem for the heavy quarks, we clearly ignore the quantum processes and in particular
the hard gluon emission by bremsstrahlung: these processes might be introduced
by adding a random force in the equations of motion of the heavy quarks,
but it is known that these would contribute only in a range of $p_T$ way larger than the one
that we consider here.

We also study in detail the momentum broadening, $\sigma_p=\langle (p(t)-p_0)^2\rangle$, 
where $p_0$ denotes the initial value of momentum of charm. 
For a standard Brownian motion with uncorrelated noise $\sigma_p = 2{\cal D} t$ for $t\ll1/\gamma$,
while the drag affects later evolution.
For charm we find that $\sigma_p\propto t^2$ in the very early part of the evolution:
we interpret this as the effect of the memory in the correlator of the force exerted by the gluon fields on the charm;
in fact, for a Brownian motion with a nontrivial memory kernel $\sigma_p\propto t^2$.
In the case of charm in the Ev-Glasma this can be interpreted as a time correlation of the force, $\bm F$,
acting on the charm, 
$\langle \bm F(\bm x(\tau_1),\tau_1)\bm F(\bm x(\tau_2),\tau_2)\rangle \neq A\delta(\tau_1-\tau_2)$,
where $\bm x(\tau)$ denotes the position of the charm at time $\tau_1$.
Although we do not compute the correlator of the force since it would require a different approach to the
solution of the CYM equations, we have computed the correlator
of the electric field at different times and found that this is characterized by a finite time decay,
suggesting finite time correlation of the force.
Comparison with the $\sigma_p$ of a standard, uncorrelated Brownian motion we find that 
the effect of memory is to slow down the diffusion in the very early stage, but after a short transient
$\sigma_p$ in the Ev-Glasma overshoots the one of the uncorrelated motion.

We complete the study by computing diffusion in the transverse coordinate space. 
We find that $\sigma_x=\langle (x(t)-x_0)^2\rangle$
follows the qualitative path of a Brownian motion in its early stage, $\sigma_x = a t^2 + b t^3$.
In this stage memory plays a less relevant role for $\sigma_x$ because it would affect only terms of order $O(t^4)$
which are smaller than the $O(t^2)$.
The $\sigma_x\propto t^2$ shows that the diffusion of charm in Ev-Glasma is 
in a ballistic regime \cite{levywalk,Zaslavsky}.

The plan of the article is as follows: in Section~\ref{Sec:glasma} we review the theoretical setup of the work;
in Section~\ref{sec:quick} we review briefly the solution of the Langevin equations for the Brownian motion,
emphasizing the effect of a memory kernel on the early evolution of $\sigma_p$;
in Section~\ref{Sec:bqwe} we present our results on $\sigma_p$, $\sigma_x$ and $R_\mathrm{AA}$ of charm;
in Section~\ref{Sec:MC} we compare $\sigma_p$ of charm in the Ev-Glasma and in a thermal medium
and present a qualitative comparison of the $R_\mathrm{AA}$ in the two cases; finally in Section~\ref{Sec:kli}
we summarize our work and discuss possible future improvements.

\section{The model\label{Sec:glasma}}

\subsection{Glasma and classical Yang-Mills equations}

In this section, we review the Glasma and the McLerran-Venugopalan (MV) 
model \cite{McLerran:1993ni,McLerran:1993ka,McLerran:1994vd,Kovchegov:1996ty}. 
In this work we scale the gauge fields as 
$A_\mu \rightarrow A_\mu/g$ where $g$ is the QCD coupling.
In the MV model for the collision of two nuclei labeled as $A$ and $B$ the static color charge densities $\rho_a$   
on $A$ and $B$ are assumed to be random variables that are 
normally distributed with zero mean 
and variance given by  
\begin{equation}
\langle \rho^a_{A,B}(\bm x_T)\rho^b_{A,B}(\bm y_T)\rangle = 
(g^2\mu_{A,B})^2  \delta^{ab}\delta^{(2)}(\bm x_T-\bm y_T);
\label{eq:dfg}
\end{equation}
here,  $a$ and $b$ denote the adjoint color index;
in this work we consider the case of the $SU(2)$ color group therefore
$a,b=1,2,3$.  The choice of $SU(2)$ rather than $SU(3)$ is done for simplicity,
because it allows to implement easily the equations of motion and the initialization of the gauge fields
using linear representations of the exponential operators; an upgrade of our code to the $SU(3)$ case is a work in progress
and results will appear in the near future.
In Eq.~(\ref{eq:dfg}) $g^2\mu_A$ denotes the color charge density,  
 $g^2\mu=O(Q_s)$ \cite{Lappi:2007ku}.
For protons, the dependence of $Q_s$ of the average $x=\langle p_T\rangle/\sqrt{s}$ 
can be estimated via the GBW fit  \cite{GolecBiernat:1999qd,GolecBiernat:1998js,Kovchegov:2012mbw,Albacete:2012xq},
$
Q_s^2 = Q_{0}^2 \left(x_0/x \right)^\lambda
$,
with $\lambda=0.277$, $Q_0=1$ GeV and $x_0=4.1\times 10^{-5}$.
For nuclei we borrow the modification of the GBW fit obtained within the IP-Sat model \cite{Kowalski:2007rw},
namely
\begin{equation}
Q_s^2 =c A^{1/3}\log AQ_{s,0}^2 \left(\frac{x_0}{x}\right)^\lambda.\label{eq:QS_slide2}
\end{equation}
Other forms of the generalized GBW fit are 
possible  \cite{Armesto:2004ud,Freund:2002ux}, but these do not lead to significant changes of $Q_s$.
Using the numerical result $Q_s/g^2\mu=0.57$ of  \cite{Lappi:2007ku} we find 
$g^2\mu_\mathrm{Pb} = 3.4$ GeV for the Pb nucleus for collisions at the LHC energy,
or $Q_s=1.9$ GeV.

The static color sources $\{\rho\}$ generate pure gauge fields outside and on the light cone, which in the forward light cone combine 
and give the initial Glasma fields. In order to determine these fields
we solve the Poisson equations for the gauge potentials
generated by  $\rho_A$ and $\rho_B$, namely
\begin{equation}
-\partial_\perp^2 \Lambda^{(A,B)}(\bm x_T) = \rho^{(A,B)}(\bm x_T).
\end{equation}
Wilson lines are computed as
$
V^\dagger(\bm x_T) = e^{i \Lambda^{(A)}(\bm x_T)}$, 
$W^\dagger(\bm x_T) = e^{i \Lambda^{(B)}(\bm x_T)}$,
and the pure gauge fields of the two colliding nuclei are given by
$
\alpha_i^{(A)} = i V \partial_i V^\dagger$,
$\alpha_i^{(B)} = i W \partial_i W^\dagger$.
In terms of these fields the solution of the CYM in the forward light cone
at initial time, namely the Glasma gauge potential, 
can be written as 
$A_i = \alpha_i^{(A)} + \alpha_i^{(B)}$~ for $i=x,y$ and $A_z = 0$,
and the Glasma fields are \cite{Kovner:1995ja,Kovner:1995ts} 
\begin{eqnarray}
&& E^z = i\sum_{i=x,y}\left[\alpha_i^{(B)},\alpha_i^{(A)}\right], \label{eq:f1}\\
&& B^z = i\left(
\left[\alpha_x^{(B)},\alpha_y^{(A)}\right]  + \left[\alpha_x^{(A)},\alpha_y^{(B)}\right]  
\right),\label{eq:f2}
\end{eqnarray}
where $z$ is the direction of the collision and the transverse fields vanish.
 
The evolution of the initial condition is achieved via the 
classical Yang-Mills (CYM) equations, namely 
\begin{eqnarray}
\frac{dA_i^a(x)}{dt} &=& E_i^a(x),\\
\frac{dE_i^a(x)}{dt} &=&   \partial_j F_{ji}^a(x) + 
  f^{abc} A_j^b(x)  F_{ji}^c(x) - j_i^a(x);\label{eq:CYM_el}
\end{eqnarray}
we have put
\begin{equation}
F_{ij}^a(x) = \partial_i A_j^a(x) - \partial_j A_i^a(x)  + f^{abc} A_i^b(x) A_j^c(x),
\label{eq:Fij}
\end{equation}
where $f^{abc} = \varepsilon^{abc}$ with $\varepsilon^{123} = +1$, and the standard summation convention has been used.
We name the evolving field as the Ev-Glasma, leaving the name Glasma to the initial condition.
At variance with previous calculations 
\cite{Liu:2019lac,Sun:2019fud,Ruggieri:2018rzi,Ipp:2020nfu,Boguslavski:2020tqz,Mrowczynski:2017kso,Carrington:2020sww}
we include the color current, $ j_i^a$, carried by the heavy quarks. 
This is essential to describe the energy loss of the colored particles interacting with the evolving 
Glasma \cite{Jamal:2020fxo,Han:2017nfz,Jamal:2019svc,Jamal:2020emj}, due to the polarization of the medium.

The lack $j_i^a$ in previous calculations gives a purely diffusive motion of heavy quarks,
and interaction with the gluon fields leads merely to momentum broadening.
Instead, adding the color current and solving consistently the field equations and the kinetic equations of the heavy quarks,
see below, we take into account both momentum broadening and energy loss. 
Our solution of the problem is purely numerical, however we do not rely on any assumption on
equilibration and thermalization of both the gluon medium and the heavy quarks,
as well as we do not assume linear response theory and do not make any assumption on the 
trajectories and velocities of the heavy quarks.
This approach solves the problem of propagation of heavy quarks in the Ev-Glasma
completely, as far as classical field theory can do; this solution is similar to Electrodynamics,
in which solving consistently the Maxwell equations with the kinetic equations for the charges
gives both near and far fields produced by the charges themselves; in particular,
the far fields are responsible of electromagnetic radiation.

\subsection{Wong equations for heavy quarks}
The dynamics of charm quarks in the Ev-Glasma is studied by the Wong equations  
\cite{Wong:1970fu,Boozer,Liu:2019lac,Sun:2019fud,Ruggieri:2018rzi,Ipp:2020nfu}, that
for a single quark can be written as
\begin{eqnarray}
&&\frac{d x_i}{dt} = \frac{p_i}{E},\\
&&E\frac{d p_i}{dt} = Q_a F^a_{i\nu}p^\nu,\\
&&E\frac{d Q_a}{dt} = - Q_c\varepsilon^{cba} \bm A_b\cdot\bm p,\label{eq:11}
\end{eqnarray}
where $i=x,y,z$; these correspond to the Hamilton equations of motion for the coordinate and its conjugate
momentum, while  the third equation corresponds to a classical description of the conservation of the color current.
Here $E = \sqrt{\bm p^2 + m^2}$ with $m=m_c=1.5$ GeV.  

In the third Wong equations, $Q_a$ with $a=1,\dots,N_c^2-1$ corresponds to an effective 
color charge of quarks;
this has not be confused with the standard QCD color charge of quarks,
because quarks sit in the fundamental representation of the color group $SU(N_c)$
so they carry $N_c$ colors, while $Q^a$ has the adjoint color index.
This effective charge can be understood as a function that allows to describe classically the color current
carried by the heavy quarks, namely \cite{Ryblewski:2013eja,Ruggieri:2015yea,Oliva:2017pri} 
\begin{equation}
j_i^a(x)= g^2 Q_a\int \frac{d^3p}{E}p_i f_a(p,x)
\end{equation}
in the continuum limit, or 
\begin{equation}
j_i^a=g^2\sum Q_a p_i/E,\label{eq:cam}
\end{equation}
in the discretized version, where the sum is understood over all 
particles present in a given lattice cell. 
Notice that the current is multiplied by the squared of the QCD coupling, $g^2$:
this is simply due to the scaling of the gluon fields used for the CYM equations mentioned  
in Section~\ref{Sec:glasma} and can be verified easily starting from the QCD lagrangian.
 For each heavy quark the set of $Q_a$ is initialized with uniform probability on the sphere
 $Q^2=Q_1^2 + Q_2^2 + Q_3^2=1$; this is achieved by extracting
 a random
  number, $z$, with uniform probability in the range $(-1,1)$ which represents
  $z=\cos\theta$ with $\theta$ the polar angle, and another random
  number, $\phi$, with uniform probability in the range $(0,2\pi)$ representing the azimuthal angle. Then, we put
  \begin{eqnarray}
  Q_1 &=& \cos\phi\sqrt{1-z^2},\\
  Q_2 &=& \sin\phi\sqrt{1-z^2},\\
  Q_3 &=& z.
  \end{eqnarray}
Note that $Q^2$ is constant in  the evolution. 
This can be proved by multiplying both sides of Eq.~\eqref{eq:11} by $Q_a$ and summing over $a=1,2,3$:
  on the left hand side we would have a term proportional to $E~d Q^2/dt$,
 while on the right hand side we would have 
 a term proportional to $\varepsilon_{abc} Q_c Q_a$: this vanihes because
 the antisymmetric $\varepsilon_{abc}$ is saturated with the symmetric
 tensor $Q_a Q_c$. Therefore, $d Q^2/dt=0$ for each heavy quark:
 the interaction of the heavy quarks with the gluon field gives kicks to the color charge vector
 $(Q_1,Q_2,Q_3)$ but does not change its magnitude.
For each heavy quark we produce an antiquark as well: for this, we assume the same
initial position of the companion quark, opposite momentum and a random color charge. 
 Solving the Wong equations is equivalent to solve the Boltzmann-Vlasov equations for a collisionless plasma
made of heavy quarks, which propagate in the Ev-Glasma
We enlarge the number of heavy quarks by $N_p$ test particles to improve statistics: 
this amounts to replace $g^2\rightarrow g^2/N_p$ in Eq.~\eqref{eq:cam}.

Heavy quarks are initialized at time $\tau_\mathrm{form}=1/(2m)$.
In calculations based on relativistic transport the heavy quarks are assumed to do a free streaming between their formation
time and the initialization of the quark-gluon plasma phase, see \cite{Scardina:2017ipo} 
and references therein;
we do not have a free streaming period and the heavy quarks are formed exactly at their formation time
and interact immediately with the gluon background.

\section{A quick reminder on the diffusion in the Brownian motion\label{sec:quick}}     
In this section, we review briefly the classical Brownian motion in one spatial dimension:
this reminder is useful to fix a few key results, that allow to understand
better those that we obtain for the motion of the heavy quarks in the Ev-Glasma.
For simplicity we study only the case of a nonrelativistic particle: results about momentum broadening
are valid also in the relativistic case. In order to highlight the most important characteristics of the Brownian motion
we make several assumptions along the way: these assumptions have illustrative purposes and are not done
in the full solution of the problem for charm  presented in the next section. 
    
Brownian motion is the motion of a heavy particle, with mass $M$,
in a bath that interacts with the particle via 
a time-dependent random force, $\xi(t)$, plus a viscous force, $f_\mathrm{drag}$ given by
\begin{equation}
f_\mathrm{drag}=-\int_0^t\gamma(t-t_1) p(t_1)dt_1. \label{eq:withdrag}
\end{equation}
 Here $\gamma(t)$ is the dissipative kernel, that in general can depend on time.
The evolution of momentum of the heavy particle is governed by the equation
\begin{eqnarray}
\frac{dp}{dt} &=& -\int_0^t\gamma(t-t_1) p(t_1)dt_1 + \xi(t).\label{eq:gn3}
\end{eqnarray}
In this equation, $f_\mathrm{drag}$ is responsible of the energy loss of the heavy particle, while
$\xi(t)$ causes momentum broadening, see also the discussion below.
In particular,
the viscous force is necessary for the equilibration of the  heavy particle with the medium.

The random force $\xi$ is assumed to  satisfy $\langle\xi(t)\rangle =0$ and
\begin{equation}
\langle\xi(t_1)\xi(t_2)\rangle = 2{\cal D} f(t_1-t_2);\label{eq:gn1mem}
\end{equation}
in the simplest case, the time correlations of the random force are neglected so it is assumed that 
\begin{equation}
f(t_1 - t_2) = \delta(t_1-t_2),\label{eq:gn1}
\end{equation}
namely
the motion of the heavy particle is a Markov process.
On the other hand, for the very early stage of the propagation of heavy quarks in the Ev-Glasma 
it is useful to introduce memory effects, namely assume that the time correlator of the random force
driven by the gluon fields
vanishes only if $|t_1-t_2|\gg \tau_\mathrm{mem}$.

For the discussion we assume $\tau_\mathrm{mem}\ll\tau_\mathrm{therm}$,  which we have verified 
{\it a posteriori} to be  satisfied by heavy quarks
in the Ev-Glasma fields. Under this assumption, it is legitimate to consider Eq.~\eqref{eq:gn3} in three limits,
namely $t\ll\tau_\mathrm{mem}$ that we call the very early stage, 
$\tau_\mathrm{mem}\ll t \ll \tau_\mathrm{therm}$ that we call the pre-equilibrium stage 
and $\tau_\mathrm{therm}\ll t$ that we call the equilibrium stage.

Firstly, we focus on the very early stage.
The general solution of Eq.~\eqref{eq:gn3} can be written in terms of Laplace transforms \cite{Ruggieri:2019zos}
and will be presented elsewhere;
for the purpose of the present study it suffices to say that
for $t\ll\tau_\mathrm{mem}\ll\tau_\mathrm{therm}$ the drag force can be neglected 
and, putting  $\sigma_p\equiv \langle (p(t)-P_0)^2\rangle$ with $P_0 = p(t=0)$ we have 
\begin{equation}
\sigma_p = 2{\cal D} \int_0^t\! dt_1\!\!\int_0^t\! dt_2~f(t_1-t_2)
\equiv 2 {\cal D}F(t),\label{eq:wehave1}
\end{equation}
where we have used $\langle\xi(t)\rangle=0$,
$\langle P_0 \xi(t)\rangle = 0$ as well as Eq.~\eqref{eq:gn1mem}. 
In the very early stage we can expand the right hand side of the above equation in powers of $t/\tau_\mathrm{mem}$,
namely $F(t)\approx F^\prime(0) t + F^{\prime\prime}(0)t^2/2$ with each prime denoting a derivative with respect to $t$, and 
\begin{eqnarray}
F^\prime(t) &=& 2\int_0^t\! dt_1~f(t-t_1),\\
\frac{F^{\prime\prime}(t)}{2} &=& f(0) + \lim_{t\rightarrow 0^+} \int_0^t\! dt_1~\frac{\partial}{\partial t}f(t-t_1).
\end{eqnarray}
We notice that 
if $f(t-t_1)$ has not singularities then 
$F^\prime(0)=0$. The only case in which $F^\prime(0)\neq 0$ is when
$f(t-t_1)$ is singular, for example for $t_1=t$: this happens in particular for the Markov processes. 
For the case of heavy quarks in Ev-Glasma the random force 
is related to the correlators of the electric and magnetic color fields that are not singular \cite{Mrowczynski:2017kso}
so it is meaningful to study this case.

For the sake of illustration we {\em assume}  
\begin{equation}
f (t)\equiv \frac{1}{2\tau_\mathrm{mem}}
\exp\left(-\frac{|t|}{\tau_\mathrm{mem}}\right),
\label{eq:memas}
\end{equation}
normalized as $\int_{-\infty}^{+\infty} f(t)dt=1$.
Specializing to Eq.~\eqref{eq:memas} we get $F^\prime(0)=0$ and
\begin{equation}
\sigma_p = {\cal D}\frac{t^2}{\tau_\mathrm{mem}},~~~t\ll\tau_\mathrm{mem}.
\label{eq:wehave333a}
\end{equation}
We notice that $\sigma_p\propto t^2$ in the very early stage. Clearly, changing Eq.~\eqref{eq:memas}
will not change Eq.~\eqref{eq:wehave333a} modulo an overall constant factor as long as the correlator of the force
is a regular function. Assuming such regular correlator we can write, in general,
\begin{equation}
\sigma_p =  {\cal D} F^{\prime\prime}(0) t^2.~~~t\ll\tau_\mathrm{mem}.
\label{eq:wehave333}
\end{equation}
We will use the above result in Section~\ref{Sec:kl}.

For the pre-thermalization and equilibrium stages the time range is much larger than the decay time of the memory
and we can effectively assume Eq.~\eqref{eq:gn1} instead of Eq.~\eqref{eq:memas} to evaluate $\sigma_p$;
moreover,  in this case, assuming for simplicity  $\gamma(t)=2\gamma\delta(t)$,
Eq.~\eqref{eq:gn3} is replaced by
\begin{equation}
\frac{dp}{dt} = - \gamma p  + \xi(t).\label{eq:gn3aoth}
\end{equation}
The solution of Eq.~\eqref{eq:gn3} is given by
\begin{equation}
p(t) = P_0 e^{-\gamma t} + e^{-\gamma t} \int_0^t dt_1~e^{\gamma t_1}\xi(t_1),\label{eq:gn4aaa}
\end{equation}
where $P_0 = p(t=0)$.
After a straightforward calculation we get
\begin{equation}
\sigma_p= P_0^2 (e^{-\gamma t}-1)^2 +  \frac{{\cal D}}{\gamma} (1- e^{-2\gamma t}), 
\label{eq:gn5}
\end{equation}
which represents the standard momentum broadening of a particle subject to a Brownian motion without memory.
From the above equation we get, in the pre-thermalization and equilibrium stages,
\begin{equation}
\sigma_p \approx2  {\cal D} t + \gamma(P_0^2\gamma-{\cal D}) t^2,
~~~t/\tau_\mathrm{therm}\ll 1, \label{eq:gn6}
\end{equation}
and
\begin{equation}
\sigma_p \asymp \frac{ {\cal D}}{\gamma} + P_0^2,~~~t/\tau_\mathrm{therm}\gg 1,\label{eq:gn7}
\end{equation}
where $\asymp$ means that the quantity on the left tends asymptotically to the one on the right in the large time limit.
 Notice that although we consider
the nonrelativistic limit in this Section,
Eqs.~\eqref{eq:gn6} and~\eqref{eq:gn7} are valid {\it also} in the relativistic limit: in fact, no assumption
has been done on the relation between energy and momentum.
In particular, Eq.~\eqref{eq:gn7}  implies that $\langle p^2(t)\rangle \asymp {\cal D}/\gamma$. 
It can easily be verified that the evolution of $\sigma_p$ at small and large times
in Eqs.~\eqref{eq:gn6} and~\eqref{eq:gn7} agrees with that obtained
by the solution of the one-dimensional Fokker-Planck equation solved with a $\delta-$function
initial condition \cite{Svetitsky:1987gq}.

The physical meaning of Eqs.~\eqref{eq:wehave333},~\eqref{eq:gn6} and~\eqref{eq:gn7} is that 
for times much smaller than the memory time, momentum spreads quadratically with time
until the heavy particle enters the pre-thermalization stage,
with linear momentum spreading. In both these regimes the motion is dominated by the diffusion,
while the drag force appears only to higher orders in time and affects the later motion of the heavy particle.
On the other hand, for times much larger than the thermalization time,
the heavy particle equilibrates with the medium, as a resut
of the balance of the momentum spreading given by $\xi(t)$ and the energy loss driven by $f_\mathrm{drag}$.

From $dx/dt=p/M$ we get, 
putting $\sigma_x \equiv\langle (x(t)-x_0)^2\rangle$, 
\begin{eqnarray}
\sigma_x  &=& \frac{P_0^2}{M^2}\frac{(1-e^{-\gamma t})^2}{\gamma^2}\nonumber\\
&&+\frac{2 {\cal D}}{M^2} \frac{4e^{-\gamma t} -e^{-2\gamma t}-3}{2\gamma^3} + \frac{2 {\cal D}}{M^2\gamma^2} t;
\end{eqnarray}
 we notice the last addendum on the right hand side of the above equation, that gives the characteristic
$\sigma_x\propto t$ at large times.
Once again, it is convenient to consider the limits of small and large times; for the former we get
\begin{equation}
\sigma_x\approx \frac{P_0^2}{M^2}t^2 + \frac{2 {\cal D}-3P_0^2\gamma}{3M^2}t^3,~~~t/\tau_\mathrm{therm}\ll 1, \label{eq:gn9}
\end{equation}
and for the large times
\begin{equation}
\sigma_x\asymp\frac{2 {\cal D}}{M^2\gamma^2}t,~~~t/\tau_\mathrm{therm}\gg 1, \label{eq:gn10}
\end{equation}
which is the classic result of the Brownian motion. 
Considering the very early stage in which memory is effective adds a term $O(t^4)$ to Eq.~\eqref{eq:gn9}
which is less important than the ballistic term $O(t^2)$.
In plain words, Eqs.~\eqref{eq:gn9} and~\eqref{eq:gn10} state that
for times much smaller than the thermalization time the heavy particle experiences an accelerated motion due to the
random force; this motion is gradually slowed down by the friction and waiting enough time,
the balance between friction and random force will lead to the linear spreading of the position of the particle.

\section{Results\label{Sec:bqwe} }

\subsection{Diffusion  in the transverse momentum space: toy model initial condition\label{Sec:kl}}

\begin{figure}[t!]
\begin{center}
\includegraphics[scale=0.2]{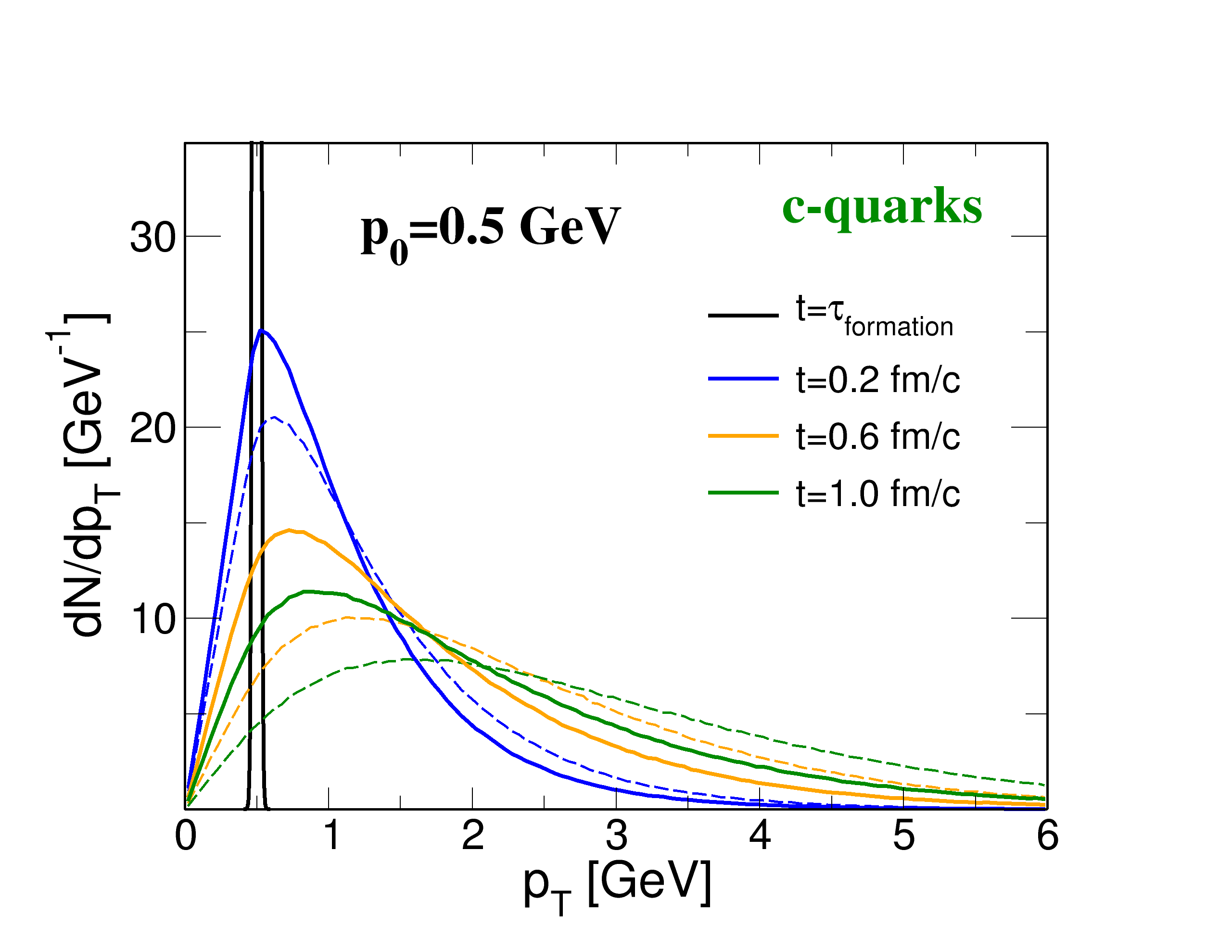}\\
\includegraphics[scale=0.2]{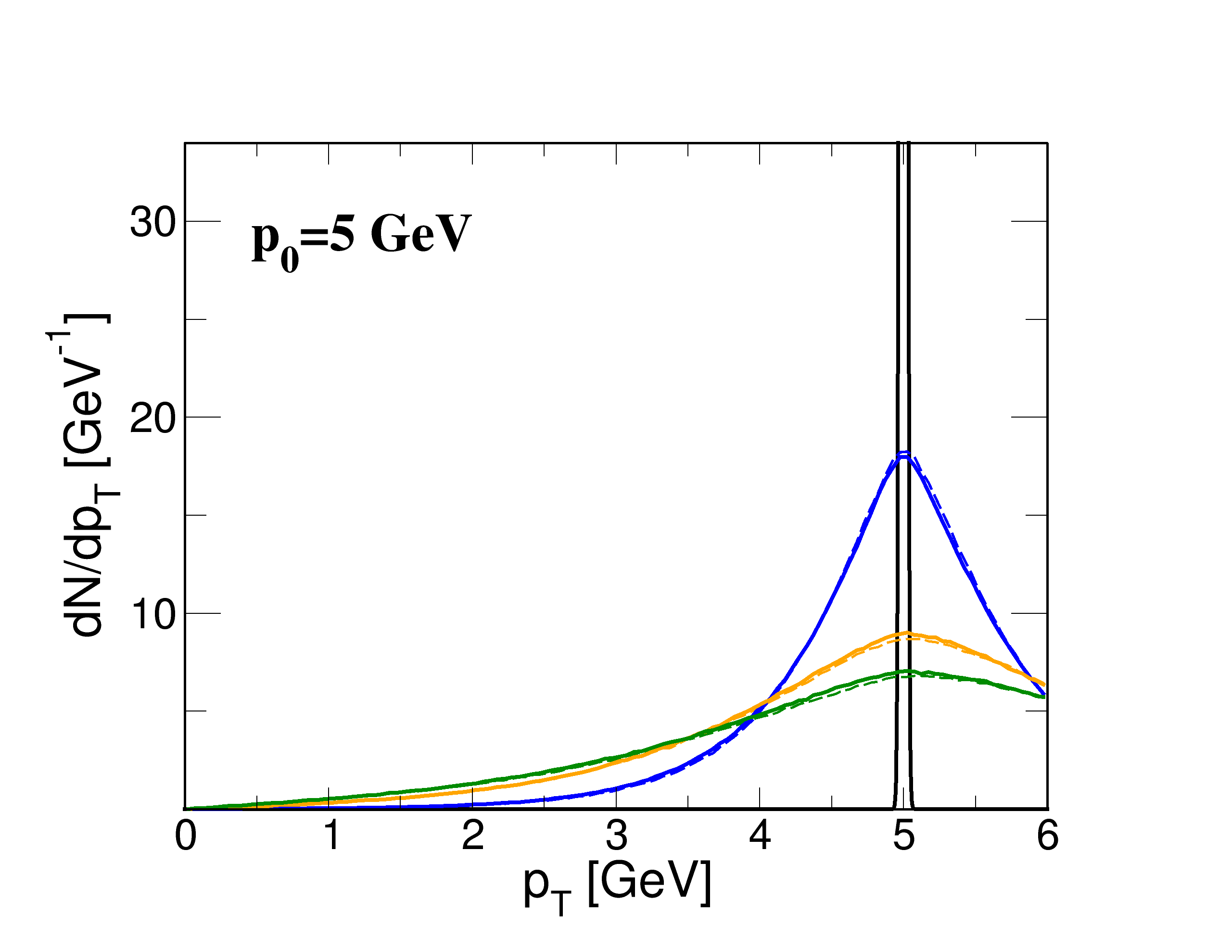}
\end{center}
\caption{\label{Fig:3a}
Color online. Distribution function,
at $t=0.2$ fm/c (blue lines), $t=0.6$ fm/c (orange lines)
and at $t=1$ fm/c (green lines); the solid lines correspond to calculations with the color current
while the dashed line to those without the current. Initial momentum is $p_0=0.5$ GeV in the upper panel
and $p_0=5$ GeV in the lower panel. Calculations correspond to $\alpha_s=0.3$.}
\end{figure}

To begin with, we prepare a $\delta-$like initialization in transverse momentum, $p_T$, and study the evolution 
of the distribution function, $dN/dp_T$,
and of momentum and energy of the charm quark. 
For this toy model initalization we use $n_c=15$ heavy quarks,
that roughly corresponds to the number of charm quarks produced in Pb-Pb collisions at 
midrapidity at the LHC energies \cite{Plumari:2017ntm}.
We show the results for charm quarks only, since they look very similar for the case of beauty quarks.

 In Fig.~\ref{Fig:3a} we plot
$dN/dp_T$ at $t=0.2$ fm/c (blue lines), $t=0.6$ fm/c (orange lines)
and at $t=1$ fm/c (green lines); the solid lines correspond to calculations with the color current
while the dashed line to those without the current. The upper panel corresponds for an initial $p_T=0.5$ GeV
while in the lower panel the initial $p_T=5$ GeV; in all calculations we initialize the quarks with 
momentum $p_z=0$.
The effect of the backreaction, due to the color current, on the charm quark is evident in the small $p_T$ case:
in fact, the evolution of $dN/dp_T$ when the current is taken into account is slower with respect to the case
in which the current is not considered, as expected by a drag force;
this is seen by the naked eye by examining both the evolution of the peak value 
and the broadening of  $dN/dp_T$.
When we consider larger values of $p_T$, we find that the effect of the drag force is not strong.

\begin{figure}[t!]
\begin{center}
\includegraphics[scale=0.2]{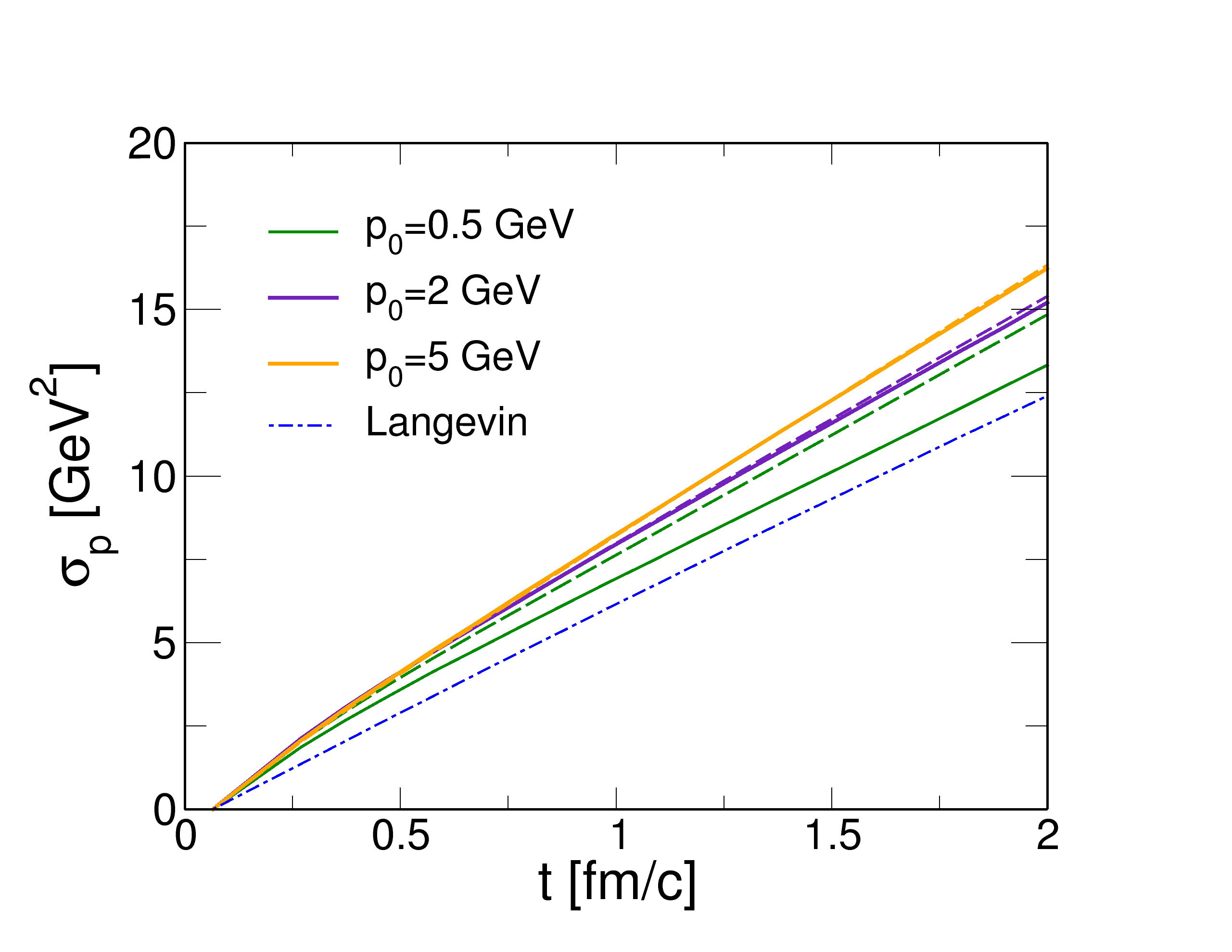}\\
\includegraphics[scale=0.2]{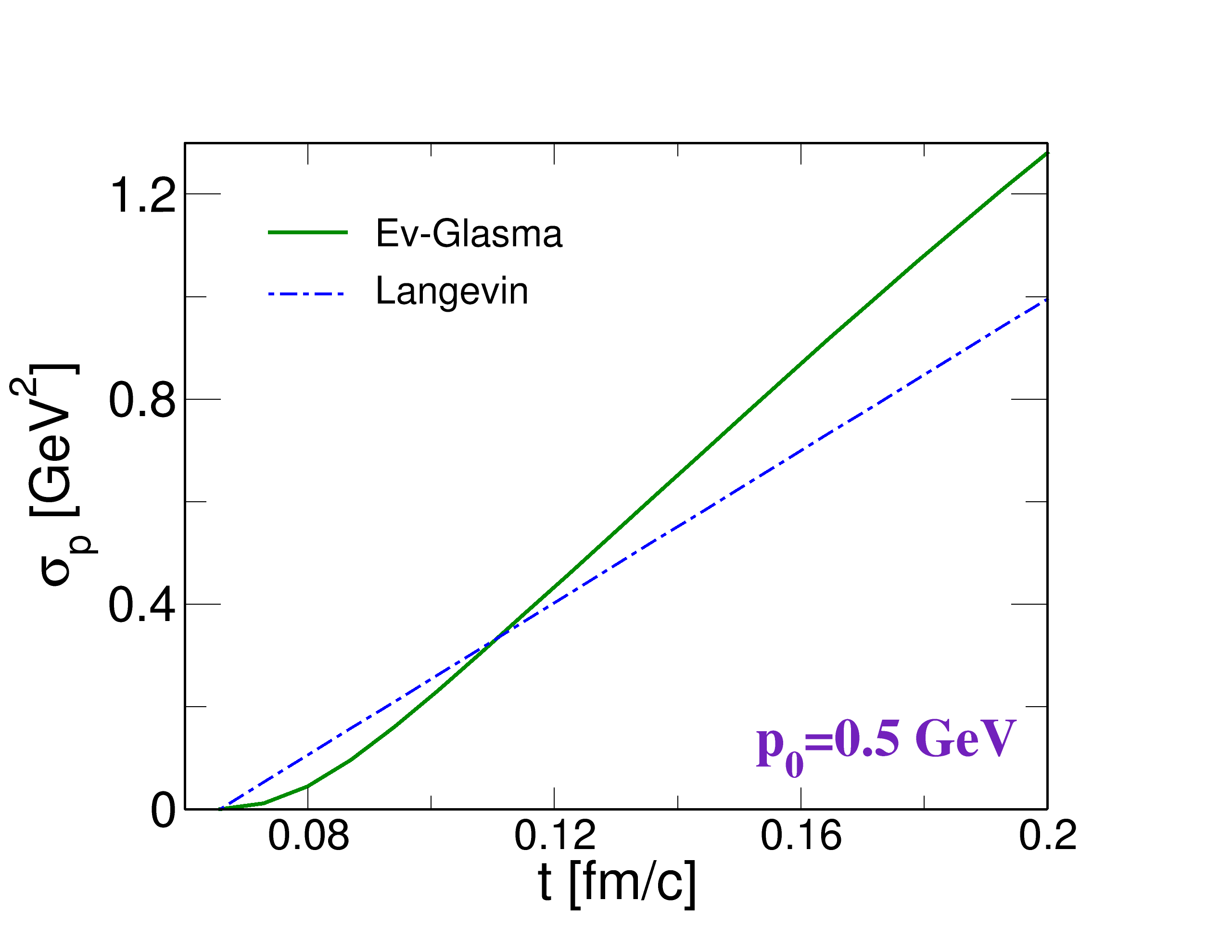}
\end{center}
\caption{\label{Fig:3aNCdelta}
Color online. Time evolution of $\sigma_p$ of charm quarks with different initial momenta.
Solid lines correspond to the calculations with the color current, dashed lines stand for the results without color current.
In the lower panel we zoom in the early time region for the case $p_0=0.5$ GeV to remark the nonlinear behavior of $\sigma_p(t)$.}
\end{figure}

In Fig.~\ref{Fig:3aNCdelta} we plot the momentum variance of charm quarks, $\sigma_p$, versus time, 
for three values of the initial $p_T$,
where we have put
\begin{equation}
\sigma_p =\frac{1}{2} \left\langle( p_x(t)-p_{0x})^2 +  ( p_y(t)-p_{0y})^2\right\rangle,\label{eq:2dim}
\end{equation}
with $p_{0x},p_{0y}$ denoting the $x,y$ components of the initial transverse momentum and $p_T^2=p_x^2 + p_y^2$.
The solid lines in Fig.~\ref{Fig:3aNCdelta} denote the 
results with current taken in to account, dashed lines correspond to calculations without current.
Results correspond to $g^2\mu=3.4$ GeV.
We notice that at small $p_T$ the effect of the drag force is quite large, lowering the momentum broadening
of $\approx 40\%$ after $t=1$ fm/c of evolution in the gluon field; 
the effect of the current becomes smaller for larger values of $p_T$.
However, for the typical lifetime of the pre-hydro stage in nuclear collisions at the LHC energy, $\tau\approx 0.3$ 
fm/c~\cite{Plumari:2017ntm},
we  find that even for small $p_T$ including the drag force does not affect the $\sigma_p$ substantially:
for example, for $p_0=0.5$ GeV we find that the effect of the current is to lower $\sigma_p$ of $\approx 13\%$.

 We notice that  
the evolution of  $\sigma_p(t)$ is not linear in the whole time range considered, 
see in particular the early time behavior of $p_0=0.5$ GeV in the lower panel of Fig.~\ref{Fig:3aNCdelta}
which is the one more relevant for the role of Ev-Glasma in the early stage of relativistic heavy ion collisions.
This nonlinearity be understood as the effect of memory in the very early stage of the evolution of charm
in the gluon field, similarly to the early stage of the non-Markovian Brownian motion 
discussed in Section \ref{sec:quick} for which $\sigma_p\propto t^2$. 
The calculation of the correlators of the force
is beyond the purpose of this article, however to check the plausibility of 
memory effects in the Ev-Glasma we have computed 
the correlator of the electric field at different times and found  a decay time $\approx 0.06$ fm/c
$\approx 1/g^2\mu$.
Using Eq.~\eqref{eq:wehave1} to fit the data in the lower panel of  Fig.~\ref{Fig:3aNCdelta}
we estimate $\tau_\mathrm{mem}\approx 0.07$ fm/c, in agreement with the result of the correlator.

We  estimate the diffusion and drag coefficients 
of charm by fitting the data in Fig.~\ref{Fig:3aNCdelta} with Eq.~\eqref{eq:gn5}
up to $t=2$ fm/c,
starting from $t=0.2$ fm/c to remove the early stage that is dominated by the memory.
We get ${\cal D} =  3.37$ GeV$^2$/fm 
and $\gamma=0.026$ fm$^{-1}$ for $p_0=0.5$ GeV; we use $\gamma$ to estimate
the thermalization time of the charm in the Glasma, namely $\tau_\mathrm{therm}=1/\gamma\approx 38$ fm/c.

In Fig.~\ref{Fig:3aNCdelta} we compare the results with those obtained by solving the standard Langevin
equations without memory Eq.~\eqref{eq:gn5}, with the values of $D$ and $\gamma$ that we get for charm in the Ev-Glasma;
we represent the data with blue dot-dashed lines. We notice that initially the $\propto t^2$ of charm in the gluon fields
makes momentum broadening slower than the corresponding Markovian dynamics.
On the other hand, for $t  \gtrsim \tau_\mathrm{mem} $ the broadening in the gluon fields overshoots the Langevin results
and gives a faster diffusion in momentum space.
We present more comparisons with the Langevin dynamics in Section~\ref{Sec:MC}.

We can summarize our findings by writing that
if we had to follow
the diffusion of charm in the Ev-Glasma for times up to $\approx 1$fm/c, 
then this would appear largely as a standard Brownian motion with drag and diffusion,
however mostly dominated by diffusion since equilibration time is quite larger than $1$ fm/c.
On the other hand, limiting to consider the timescales $\approx 0.3$ fm/c which are those relevant
for the relativistic heavy ion collisions, the memory is qualitatively important as it
slows down the momentum broadening of the charm quarks in the very early stage, then gives a boost
and puts $\sigma_p$ above the result we would measure if the diffusion was a Markov process. 
Having added the drag force by the color current, we have found that 
the net effect of the drag  in this short time range is quite modest.

\subsection{Diffusion  in the transverse momentum space: realistic initial condition\label{sec:ri}} 

 Next we turn to a realistic initialization of heavy quarks in transverse momentum space. 
To this end, at the formation time we assume the prompt spectrum obtained
within 
Fixed Order + Next-to-Leading Log (FONLL) QCD that 
reproduces the D-mesons spectra in $pp$ collisions after fragmentation~\cite{FONLL, Cacciari:2012ny,Cacciari:2015fta} 
\begin{equation}
\left.\frac{dN}{d^2 p_T}\right|_\mathrm{prompt} = \frac{x_0}{(1 + x_{3}{p_{T}}^{x_1})^{x_2}};\label{eq:HQ_1}
\end{equation}
the parameters that we use in the calculations are $x_0=20.2837$, $x_1=1.95061$, $x_2=3.13695$ and $x_3=0.0751663$ for 
charm quarks;
the slope of the spectrum has been calibrated to a collision at $\sqrt{s}=5.02$ TeV.
We use $n_c=15$ charm quarks as in the $\delta-$function initializations,
corresponding to the estimated number of charm quarks produced in Pb-Pb collisions at 
midrapidity at the LHC energies \cite{Plumari:2017ntm}.
Moreover, we assume that the initial longitudinal momentum vanishes.
In coordinate space, for the setup of AA collisions, we simulate the most central interaction region
in which we assume that width of the random color density fluctuations, given by $g^2\mu$, is constant: as a consequence,
we assume that the probability of formation of heavy quarks in
this region is also uniform.

 We can quantify the effect of the propagation of charm quarks in the Ev-Glasma by introducing the modification factor,
$R_\mathrm{AA}$, defined as
\begin{equation}
R_\mathrm{AA} = \frac{\left(dN/d^2 p_T\right)_\mathrm{evolved}}{\left(dN/d^2 p_T\right)_\mathrm{prompt}},
\end{equation}
where the prompt spectrum is given by Eq.~\eqref{eq:HQ_1} and 
$\left(dN/d^2 p_T\right)_\mathrm{evolved}$ corresponds to the spectrum 
after the evolution in the Glasma fields: this is a time dependent quantity so in general $R_{\mathrm{AA}}$ depends on time as well.
If $R_\mathrm{AA}=1$ for all values of $p_T$ it means that the spectrum after the evolution
is the one computed from hard scatterings in pQCD; on the other hand,
$R_\mathrm{AA}\neq 1$ signals the interaction of the charm with the medium.

\begin{figure}[t!]
\begin{center}
\includegraphics[scale=0.2]{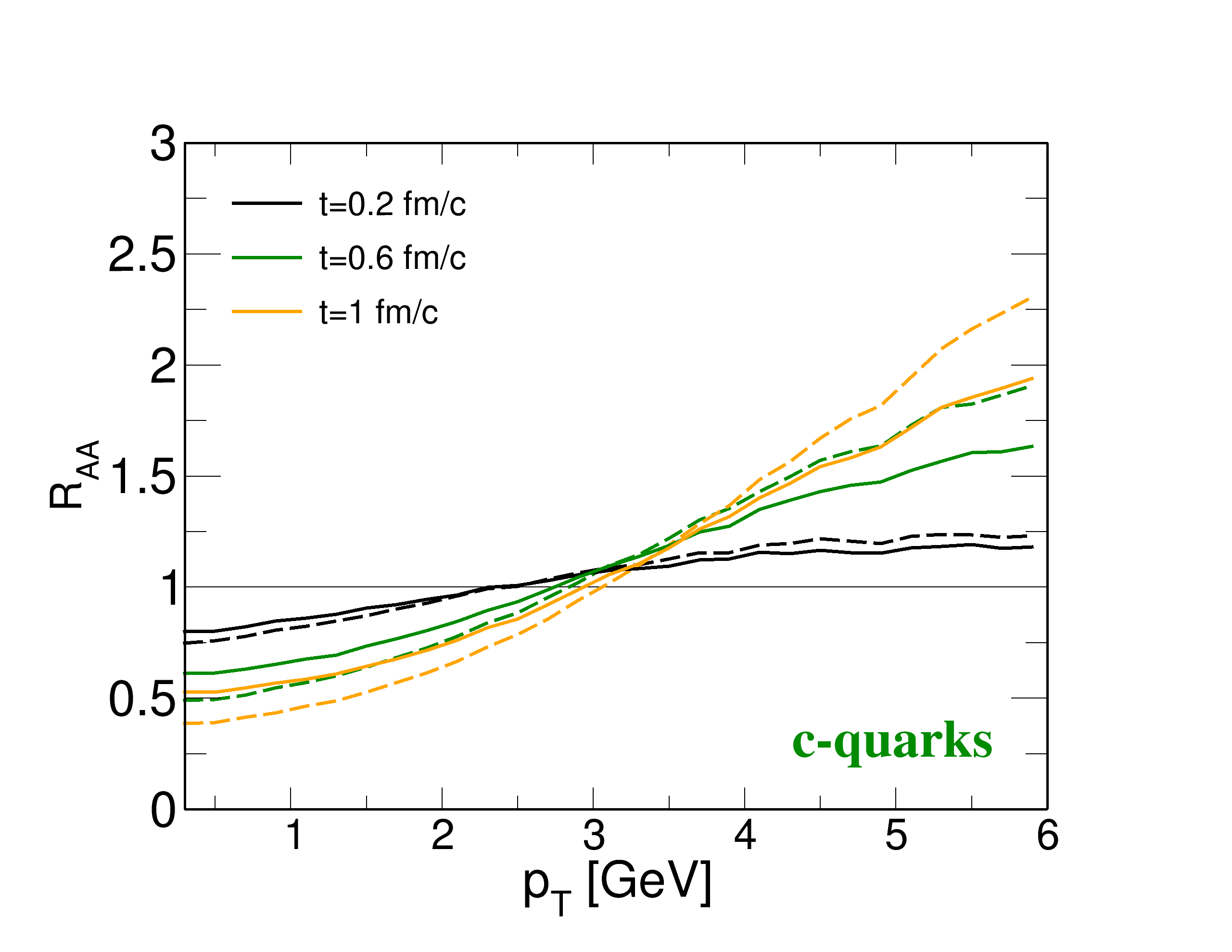}
\end{center}
\caption{\label{Fig:4}
Color online. Nuclear modification factor of charm quarks versus $p_T$, computed at different times.
Calculations with and without current are represented by solid and dashed lines respectively.}
\end{figure}

 In Fig.~\ref{Fig:4} we plot the nuclear modification factor for $c-$quarks 
at three different times, computed with and without the color current in the YM equations.
The tilting of the spectrum discussed above naturally results in $R_\mathrm{AA}$ smaller than one
at low $p_T$, as a result of the diffusion of these charms to higher $p_T$; 
the larger the time of the evolution in the gluon field, the larger the effect on $R_\mathrm{AA}$.
The drag force induced by the polarization of the medium slows down the evolution of the spectrum,
and it naturally results in a slower evolution of $R_\mathrm{AA}$ as well.

The results in Fig.~\ref{Fig:4} agree qualitatively with those presented in \cite{Liu:2019lac,Ruggieri:2018rzi}:
the main novelty of the present work is to upgrade those results taking into account the
drag force that results from the polarization of the medium induced by the color current
carried by the quarks.
Quantitatively, we find that at up to $t=0.3$ fm/c the effect of the drag force is negligible,
while it is substantial at $t=1$ fm/c.
The typical lifetime of the pre-hydro stage is $\tau\approx 0.3$ fm/c for collisions at the LHC 
energies \cite{Plumari:2017ntm}, therefore the results of the present study suggest that
the inclusion of the color current of the charm quarks will not affect drastically
the evolution of the spectrum at the LHC energies in comparison to the  results
published in \cite{Liu:2019lac,Ruggieri:2018rzi}.

The qualitative shape of $R_\mathrm{AA}$ that we get at the end of the pre-thermalization stage
is different from the one
that is usually found after the evolution in the quark-gluon plasma, see \cite{Scardina:2015caa,Das:2015ana}
and references therein:
there, it is evident the diffusion of the large $p_T$ charm quarks to lower $p_T$ states due to energy loss.
It should be noted that in the present calculation the energy density of the bulk is quite larger than the
one in the quark-gluon plasma phase, therefore in the latter the effect of the drag force will be larger
than the one we have found here. 
 In fact the energy density, $\varepsilon$, of the Ev-Glasma can be guessed to be of the order of $O(Q_s^4)$:
for $Q_s=2$ GeV, that corresponds to the value of $g^2\mu$ that we use in our simulations,
we get the educated guess $\varepsilon = O\left[ (2~\mathrm{GeV})^4\right]$; this guess, and what we compute
in the actual simulation that is $\varepsilon\approx 7$ GeV$^4$, are in the same ballpark. 
Moreover, the drag coefficient of charm in the quark-gluon plasma phase
is larger than the one we have found in the Ev-Glasma. Both these factor make the motion
of charm in the pre-thermalization stage gluon fields dominated by diffusion
and low-$p_T$ flow to higher $p_T$. 
 As a final remark, we have checked that the curves in Fig.~\ref{Fig:4} invert their tendency  
 already for $p_T\approx 7$ GeV, namely $d R_\mathrm{AA}/d p_T$ becomes
 negative and $R_\mathrm{AA} $ approaches $1$. 
 We do not show the result in Fig.~\ref{Fig:4} because it would require much more statistics
 due to the small number of charm quarks that sit in that $p_T$ region.

\subsection{Diffusion of charm in the transverse coordinate space}

\begin{figure}[t!]
\begin{center}
\includegraphics[scale=0.2]{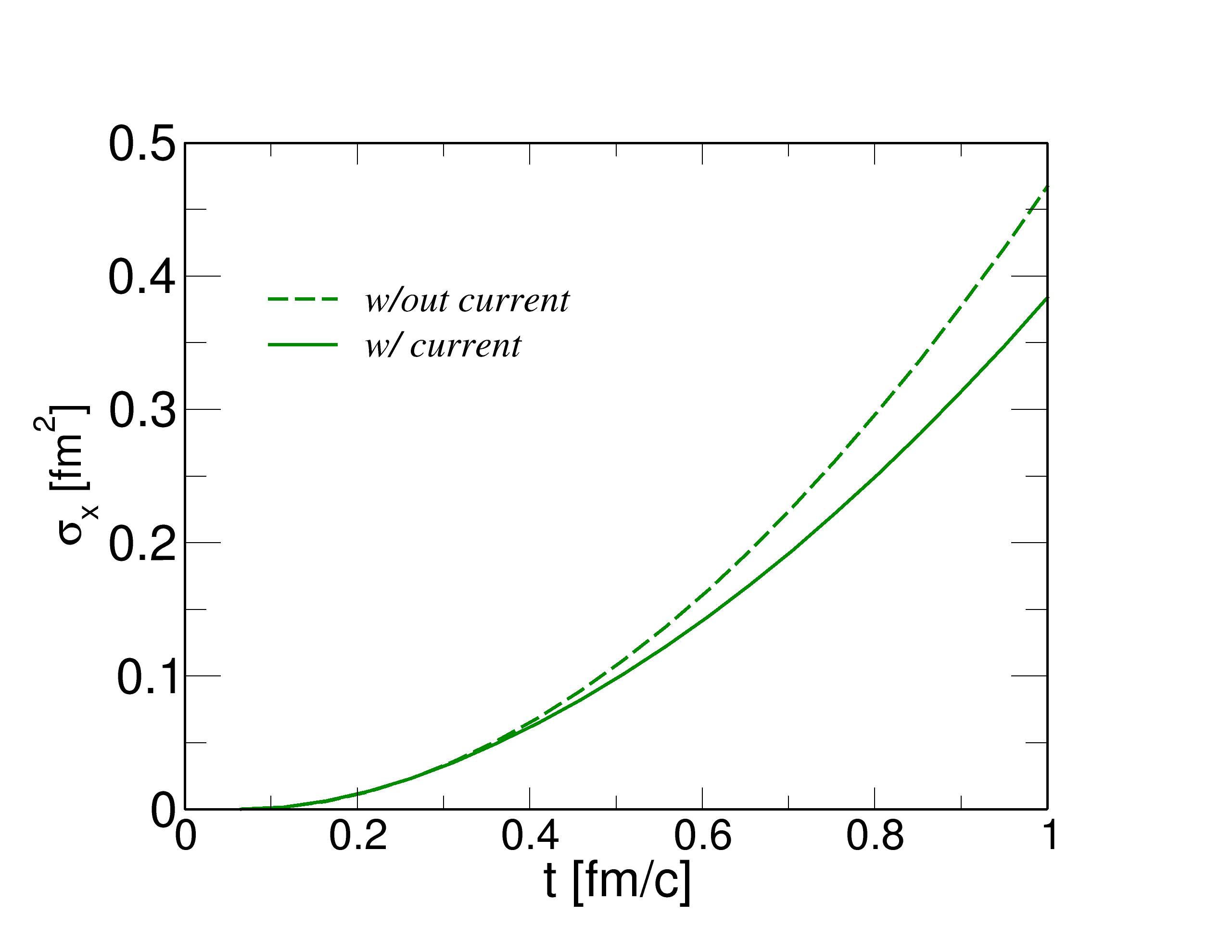}
\end{center}
\caption{\label{Fig:3aGGxx}
Color online. Transverse coordinate dispersion of charm quarks computed with and without
the color current. We take $g^2\mu = 3.4$ GeV.}
\end{figure}

In Fig.~\ref{Fig:3aGGxx} we plot the transverse coordinate variance of charm quarks, $\sigma_x$, versus time, where
\begin{equation}
\sigma_x = \left\langle( x_T(t)-x_0)^2\right\rangle,
\end{equation}
for the cases with and without the color current in the YM equations.
The calculation setup corresponds to that of Fig.~\ref{Fig:3aGGxx}. 
Energy loss slows down the diffusion since $\sigma_x$ is bent downwards when the color current is introduced. 
Once again, the effect of energy loss 
is quite modest for the very early times up to $t\approx 0.3$ fm/c, while it becomes
more substantial for larger times. 

Combining the results of this and previous subsections, we conclude that the motion of charm in the pre-thermalization
stage is that of a ballistic diffusion. In fact, equilibration time is much larger than the lifetime of the 
pre-hydro stage, which gives $\sigma_x\propto t^2$ in the time range of interest,
and  $\sigma_p\propto t^2$ in the same time range overshooting the standard diffusion for which $\sigma_p\propto t$.

\section{Comparison with Langevin dynamics\label{Sec:MC}}

In this section, we compare the evolution of charm quarks in the Ev-Glasma, with that obtained by
solving standard Langevin equations without memory. To facilitate the comparison we use the same 
diffusion coefficient in all calculations, namely ${\cal D}=3.37$ GeV$^2$/fm
that matches what we estimated in Sections~\ref{Sec:kl} and~\ref{sec:ri}.

\begin{figure}[t!]
\begin{center}
\includegraphics[scale=0.2]{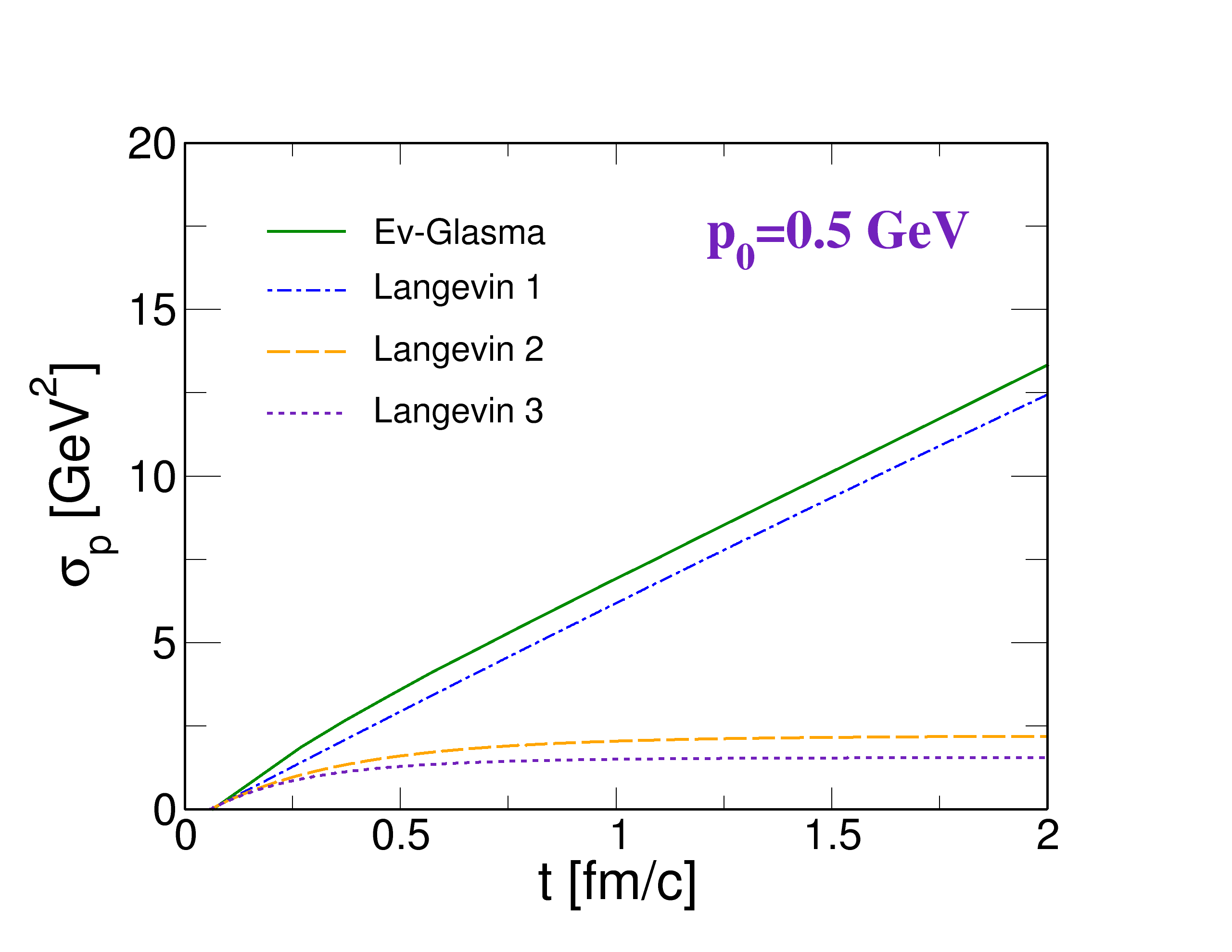}\\
\includegraphics[scale=0.2]{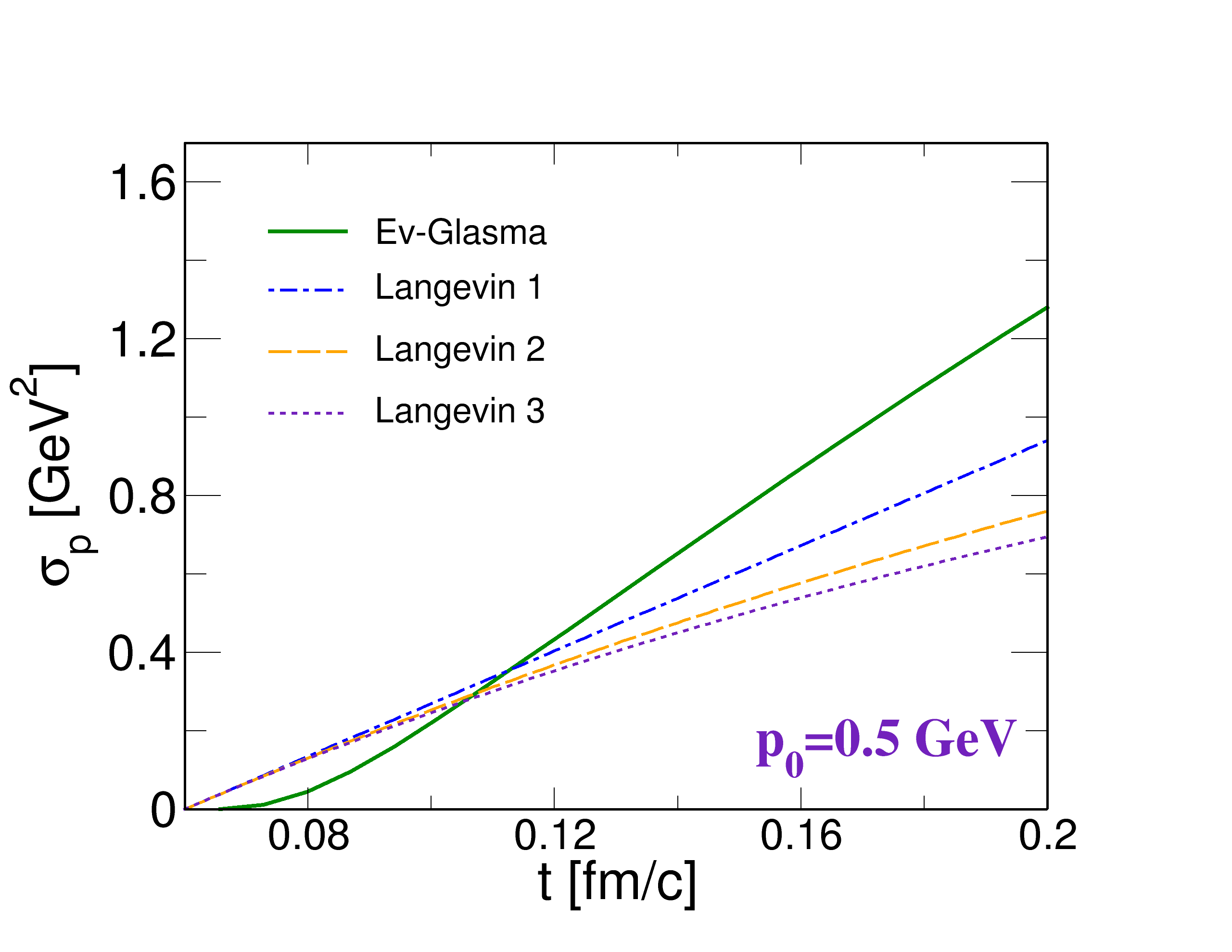}
\end{center}
\caption{\label{Fig:2acomparison}
Color online. Comparison of $\sigma_p$ of charm quarks versus time, 
between evolution in Ev-Glasma and in a thermal medium via Langevin equations.
Initialization corresponds to $p_0=0.5$ GeV. 
All Langevin calculations have the same ${\cal D}$
of Ev-Glasma, ${\cal D}=3.37$ GeV$^2$/fm.
Langevin 1 uses the same $\gamma$ of Ev-Glasma, while in 
Langevin 2 and 3 we have used the $\gamma$ that would be required by the Fluctuation-Dissipation theorem,
$\gamma={\cal D}/ET$, for $T=1.5$ GeV and $T=1$ GeV respectively: these are $\gamma=1.72$ in Langevin 2
and $\gamma=2.59$ in Langevin 3. Lower panel corresponds to a zoom up to $0.2$ fm/c.}
\end{figure}

In Fig.~\ref{Fig:2acomparison} we plot $\sigma_p$ versus time for the initialization $p_0=0.5$ GeV:
Langevin 1 uses the same $\gamma$ of Ev-Glasma namely $\gamma=0.026$ fm$^{-1}$, while in 
Langevin 2 and 3 we have used the $\gamma$ that would be required by the Fluctuation-Dissipation theorem,
$\gamma={\cal D}/ET$, for $T=1$ GeV and $T=1.5$ GeV respectively: these are $\gamma=1.72$ fm$^{-1}$ in Langevin 2
and $\gamma=2.59$ fm$^{-1}$ in Langevin 3.
In the cases Langevin 2 and 3 it is obvious that charm quarks equilibrate with the thermal medium within $\approx 1$ fm/c.
In the lower panel of Fig.~\ref{Fig:2acomparison}  we zoom on the very early stage of the evolution,
up to $t=0.2$ fm/c: we notice that $\sigma_p$ in all cases is different both qualitatively and quantitatively
from the one in the Ev-Glasma. In particular, 
the effect of memory is clearly visible in the Ev-Glasma for $\sigma_p\propto t^2$ rather than $\propto t$
up to $t\approx\tau_\mathrm{mem}$, and
$\sigma_p$ in the Ev-Glasma overshoots that in the Langevin dynamics
for $t\gtrsim\tau_\mathrm{mem}$.

\section{Conclusions and outlook\label{Sec:kli}}
We have studied the diffusion of  charm quarks in the Ev-Glasma
produced in high energy nucleus-nucleus collisions. 
We have solved consistently the Yang-Mills equations for the evolution of the gluon field
and the Wong equations for the heavy quarks. 
The main novelty of this study is the inclusion of the color
current carried by heavy quarks in the classical Yang-Mills equations of the gluon field,
that is necessary to describe the energy loss of heavy quarks due to the polarization of the 
medium \cite{Jamal:2020fxo,Han:2017nfz,Jamal:2019svc,Jamal:2020emj}.
This study concludes the one started in \cite{Ruggieri:2018rzi,Sun:2019fud,Liu:2019lac},
in which the phenomenological impact of the diffusion of heavy quarks in Glasma have been studied for the first time.
There, the idea that diffusion in the early stage affects the nuclear modification factor as well as the elliptic flow of heavy quarks,
despite the short lifetime of the pre-hydro stage, was investigated, but the calculations neglected the 
color current carried by the heavy quarks and the subsequent backreaction on the motion of heavy quarks themselves.
We fill this gap here.

Qualitatively, our results agree with the expectation that the energy loss slows down the momentum broadening 
of charm and beauty in the Ev-Glasma. This affects the evolution of the nuclear modification factor in the pre-hydro stage:
however, taking into account energy loss we get $R_\mathrm{AA}$ that is consistent, within the $10\%$, with results
previously computed without the current. This modest effect can be understood easily because
the current carried by charm and beauty is very tiny due to the low density of these quarks in the initial stage. 
Therefore, we confirm the previous findings \cite{Ruggieri:2018rzi,Sun:2019fud,Liu:2019lac} that 
the diffusion of the heavy quarks in Glasma is responsible of a tilt of their  spectrum,
effectively moving low $p_T$ quarks to higher $p_T$.
We have achieved this conclusion by studying $\delta-$function initializations as well as realistic $p_T-$initializations,
looking at both the momentum broadening and the $R_\mathrm{AA}$ of charm.

We have investigated more closely the motion of the heavy quarks.
We have found that overall the diffusion-with-drag can be interpreted in terms of the Brownian motion at late times,
plus a motion with memory effects in the very early stage of the evolution. We achieve this by studying 
the $p_T-$broadening versus time, $\sigma_p(t)$, and identify an initial range in which $\sigma_p\propto t^2$,
interpreting this as an effect of memory related to the finite time width of  the correlators of the electric and magnetic color fields.
This non-Markovian regime lasts in the very early stage of the evolution
and is dominated by diffusion,
and is followed by a standard Brownian motion regime with drag and diffusion;
the net effect of the drag is however small because the lifetime of the pre-hydro stage is smaller
than the thermalization time of heavy quarks and in this case the leading contribution to momentum broadening
comes from diffusion. 
Significant effects of drag have been found at later times, $t\approx 1$ fm/c;
however, these times are well beyond the lifetime of the pre-equilibrium stage of high energy nucleus-nucleus collisions,
therefore it cannot affect any observable. For the time range in which the Ev-Glasma can play a role in collisions
the diffusion in the early stage is the relevant one, in which the effect of memory is important.
In the very early stage we have found some quantitative difference with the diffusion of a standard Brownian motion
studied via Langevin equations without a memory kernel. In particular,
momentum broadening in the Ev-Glasma proceeds slower than the linear increase of the standard Brownian motion,
then overshoots the latter: at the end of the Ev-Glasma evolution, $\tau\approx0.4$ fm/c, 
the $\sigma_p$ that we get from Ev-Glasma is larger than the one we would obtain by Langevin dynamics
with same drag and diffusion coefficients.

We have also studied the diffusion of charm   in coordinate space. We have found that
coordinate broadening, $\sigma_x$,  evolves in the early stage as $\sigma_x\propto a t^2 + b t^3$ hence faster than the
steady state Brownian motion result $\sigma_x\propto t$.
This faster diffusion in coordinate space means that the diffusion of charm in the Ev-Glasma
is in the ballistic regime.

In conclusions, our findings support the diffusion-with-no-drag advertised in \cite{Ruggieri:2018rzi,Sun:2019fud,Liu:2019lac},
and allow to understand it  easily:
the time range relevant for the propagation of the heavy quarks in the Ev-Glasma is much smaller
than the equilibration time, and in this regime the  motion is diffusion-dominated;
a short transient where memory is important is replaced by a standard Brownian motion at later times.
The drag, that we have computed self-consistently in this study, does not affect in a considerable way
the observables that we have studied, in particular the nuclear modification factor,
because substantial energy loss is effective only on time scales comparable with the thermalization time.

While this study answers the question whether energy loss is an important ingredient to study the
diffusion of heavy quarks in the Ev-Glasma, it opens up other questions.
The role of fluctuations in the initial stage should be considered:
it is well known that fluctuations enhance isotropization already in the initial condition
of Glasma \cite{Ruggieri:2017ioa,Gelis:2013rba}, therefore it is interesting to study how 
the a larger amount of isotropization affects the evolution of the heavy quarks.
In additon to this, it is important to focus on phenomenological calculations
aimed to compute the impact of the early stage diffusion on observables,
mostly hadron spectra, two-bodies correlations and collective flows,
both in proton-nucleus and nucleus-nucleus collisions.
Protons can be initialized similarly to nuclei
according to the constituent quark model, see \cite{Schenke:2014zha,Schenke:2015aqa,
Mantysaari:2017cni,Mantysaari:2016jaz,Schlichting:2014ipa} and references therein.
We have not included the cold nuclear matter effects
\cite{Eskola:2009uj,Fujii:2013yja,Ducloue:2015gfa,Rezaeian:2012ye,
Albacete:2013ei,Albacete:2016veq,Prino:2016cni,Andronic:2015wma} in the initialization of the charm quarks,
therefore adding them is a further improvement of the present work. 
We will report on these subjects in the near future.

\begin{acknowledgements}
The authors acknowledge discussions with Gabriele Coci and with David Muller.
M. R. acknowledges John Petrucci for inspiration.
M. R. is supported by the National Science Foundation of China (Grants No.11805087 and No. 11875153)
and by the Fundamental Research Funds for the Central Universities (grant number 862946).
The work of J. H. Liu is supported by China Scholarship Council (scholarship number 201806180032).
This work is supported also from the European 
Union's Horizon 2020 research and innovation program Strong 2020 under grant agreement No. 824093.
\end{acknowledgements}

\end{document}